\documentstyle[12pt]{article}

\newcommand{\beq}{\begin{equation}}
\newcommand{\eeq}{\end{equation}}
\newcommand{\bea}{\begin{eqnarray}}
\newcommand{\eea}{\end{eqnarray}}

\newcommand{\bfx}{{\bf x}}
\newcommand{\bT}{\overline{T}}

\begin{document}

\begin{flushright}
\vbox{
\begin{tabular}{l}
	FERMILAB-PUB-97/271-T \\
	September 1997
	\end{tabular}
}

\end{flushright}
\begin{center}
\vspace{+2cm}
\Large
{\bf Multidimensional integration in a heterogeneous network
environment
} \\
\vskip 2.0cm
\large
	Sini\v{s}a Veseli \\
\vskip 0.1cm
{\small Theory Group,
Fermi National Accelerator Laboratory, P.O. Box 500, Batavia,
	\rm IL 60510}
\end{center}
\thispagestyle{empty}
\vskip 1.0cm

\begin{abstract}
We consider several issues related to the multidimensional integration
using a network of heterogeneous computers. Based on these considerations,
we develop a new general purpose scheme which can significantly
reduce the time needed for evaluation of integrals with CPU intensive
integrands. This scheme is a parallel version of the well-known adaptive
Monte Carlo method (the VEGAS algorithm), and is incorporated into a new
integration package which uses the standard set of message-passing routines
in the PVM software system.
\end{abstract}

\newpage

\section{Introduction}

Evaluation of complicated multidimensional integrals is a common
computational problem occurring in many areas of science. Calculation
of scattering amplitudes in elementary particle physics using Feynman
perturbation theory \cite{gross} is a textbook example of integrals over
four or more variables.
As   dimension of the integration volume increases, the number
of integrand evaluations required by any generalized one-dimensional
numerical method grows exponentially.
That is the major obstacle for applying any of those methods in evaluation
of multidimensional integrals.
On the other hand, the convergence rate of all Monte Carlo algorithms
is independent of the dimension of the
integral \cite{hammersley}-\cite{press}.
This property makes the Monte Carlo approach ideal for
integration over many variables.

In most applications where the functions being integrated are not
expensive in terms of the CPU time, and also in those
which do not require high statistics, or
large number of function evaluations, various Monte Carlo
algorithms usually work quite satisfactory in the sense that calculations
can be done within a  reasonable time frame.
However, high statistics integrations of CPU intensive
functions may require days, or even weeks of the CPU time on
fastest workstations presently available.\footnote{Examples of calculations
which fall into this category can be easily found in high energy
physics.}

In this note we offer a new parallel scheme which may significantly reduce
the computational time needed for Monte Carlo evaluation of multidimensional
integrals with CPU expensive integrand functions. Our approach is
based on the fact that networks connecting large numbers of
heterogeneous UNIX computers are becoming more and more widespread,
and also on the existence of several message-passing software
packages \cite{p4,pvm,mpi}, which permit  networks to be used
as single large parallel computers. Clearly, the basic idea of
performing Monte Carlo integrations using many
computers is to devise a scheme of dividing one large calculation into
a number of smaller parts, which can be handled separately and
in parallel. Nevertheless, there are many intricacies which have to be
taken into account for an efficient general purpose parallel
algorithm, suitable
for use in a heterogeneous network environment. Among
the most important issues are:
\begin{itemize}
\item[(i)] Implementation of the underlying Monte Carlo algorithm
and the random number generation.
\item[(ii)] Flexibility
to adapt to a particular network environment
and to a specific function being integrated.
\item[(iii)] Robustness with respect to occasional
failure of one or more  computers in the network.
\item[(iv)] Cost of the communication between computers participating
in the calculation.
\end{itemize}
The approach described in this paper addresses all of the above issues,
and is incorporated into a new {\em Advanced Monte Carlo Integration}
(AMCI) package. The essential ingredients of the AMCI package
are the VEGAS algorithm \cite{lepage},
and the Parallel Virtual Machine (PVM) software system \cite{pvm}.
Among various Monte Carlo schemes the VEGAS algorithm, developed
by G.P. Lepage \cite{lepage}, has shown to be one of the most efficient ones.
This highly successful general purpose algorithm has become a standard
computational tool of elementary particle physics.
On the other hand, the
PVM software system \cite{pvm} provides a unified framework within which
parallel programs can be developed in an efficient and straightforward
manner using existing computers.
Because of its simple but complete programming interface, PVM
has gained widespread acceptance in the high-performance scientific community.

The rest of the paper is organized as follows: in Section \ref{vegas}
we briefly describe the general features of Monte Carlo integration
and the VEGAS algorithm. Section \ref{mihne} contains the
discussion of parallelism issues and the description
of the parallel scheme
suitable for use in a heterogeneous network
environment. This scheme is incorporated into the AMCI package, whose
most important features  are outlined
in Section \ref{amci}.  In Section \ref{perf} we investigate
the performance of the package in various situations,
and compare it to the performance of the
ordinary VEGAS programs. Our conclusions
are given in Section \ref{conc}.

\section{Monte Carlo integration and the VEGAS algorithm}
\label{vegas}

Consider the $d$-dimensional integral of a function $f(\bfx)$, where
$\bfx = x_1,x_2,\ldots,x_d$,  over a rectangular volume $V$,
\beq
I = \int_{V} d\bfx\, f(\bfx)\ .
\label{i0}
\eeq
If $N$ points $\bfx$ are randomly selected from $V$ with
probability density $p(\bfx)$ (normalized to unity),
then it can be shown that for large $N$ the integral in Eq. (\ref{i0})
is approximated by
\beq
I\simeq S^{(1)}\ .
\label{int}
\eeq
Here, $S^{(1)}$ is defined through
\beq
S^{(k)} = \frac{1}{N} \sum_{\bfx} \left(\frac{f(\bfx)}{p(\bfx)}\right)^k\ .
\eeq
As different sets of $N$ points are chosen, the quantity $S^{(1)}$
will fluctuate about the exact value of $I$. The variance
of this fluctuation is
given by\footnote{Note that the reliable
estimates of $\sigma^2$ are possible only if the integral
$$
\int_{V} d\bfx\, \frac{f^2(\bfx)}{p(\bfx)}
$$
is finite.}
\beq
\sigma^2 \simeq \frac{S^{(2)}-(S^{(1)})^2}{N-1}\ .
\label{var}
\eeq
The standard deviation $\sigma$ indicates the accuracy of $S^{(1)}$
as an estimate of the true value of the integral.

There exist a number of methods
which can be used to
reduce the variance $\sigma^2$ for the fixed $N$. Two of the most
popular techniques are {\em importance sampling} and
{\em stratified sampling}. The first one concentrates
function evaluations where the integrand is largest in magnitude, while
the second one focuses on those regions where
the contribution to the error is largest.
However, these and other methods of variance reduction
require detailed knowledge of the integrand's behavior prior
to implementation \cite{hammersley}-\cite{press}.
Because of that, they are not appropriate for a general purpose
integration algorithm.

On the other hand,
even though the VEGAS algorithm \cite{lepage} is also primarily
based on importance sampling, the feature that  distinguishes it
from other Monte Carlo schemes is
that it is {\em adaptive} in the
sense that it {\em automatically} samples
the integrand in those regions where it is largest
in magnitude. This property makes it considerably more efficient
than non-adaptive methods in
high dimensions, or with non-analytic integrand functions.

Besides importance sampling, VEGAS  also employs
some stratified sampling, which significantly improves its
efficiency in low dimensions. When stratified sampling is used,
the algorithm divides integration volume into $M=K^{d}$ subvolumes,
where $K$ is the number of subdivisions in each of $d$ integration
dimensions.\footnote{Stratified sampling in VEGAS can be disabled.}
In all of those subvolumes VEGAS performs an
$N$-point Monte Carlo integration using importance
sampling.
Thus, the total
number of function evaluations in one iteration is given by
$N_{T}= N\times M$.

The basic idea of importance sampling in VEGAS is to construct a
multidimensional probability density function that is separable,
\beq
p(\bfx) = \prod_{i=1}^{d} p_i(x_i)\ ,
\eeq
where all $p_i$'s are normalized to unity.
The optimal one-dimensional densities for separable geometry
can be shown to be \cite{hammersley}
\beq
p_i(x_i) \propto \left[\int \left(\prod_{j\not{=}i}{dx_j
\over p_j(x_j)}\right) f^2(\bfx)\right]^{1/2}\ ,
\label{p}
\eeq
which in one dimension reduces to $p(x) \propto |f(x)|$.
The above expression immediately suggests VEGAS' adaptive
strategy: in each iteration
an $N$-point Monte Carlo integration is performed in all of $K^d$
subvolumes, using a given set of one-dimensional probability densities
(initially all constant).
Besides accumulating $S^{(1)}$ and $S^{(2)}$,
which are needed for estimating the integral
and its standard deviation,
VEGAS also accumulates $K\times d$ estimators of the right-hand side
of Eq. (\ref{p}). These are then used to determine the improved
one-dimensional densities for the next iteration.\footnote{For
details related to the refinement of the
{\em sampling grid} the reader is referred to \cite{lepage}.}
In this way, an
empirical variance reduction is gradually introduced
over several iterations, and the accuracy of integration is in general
enormously enhanced over the non-adaptive Monte Carlo methods.

For each iteration results of $M$ integrations in the different subvolumes
have to be combined to give the total integral and its variance.
We denote  $I_{i,j}$ and
$\sigma^2_{i,j}$ as results obtained for
the $j$-th subvolume and in the $i$-th iteration,
using Eqs. (\ref{int}) and (\ref{var}), respectively.
The final iteration answers for the total integral and its variance
are calculated by the relations
\bea
I_i &=& \frac{1}{M}\sum_{j=1}^{M}I_{i,j}\ ,\\
\sigma^2_i &=& \frac{1}{M^2}\sum_{j=1}^{M}\sigma^2_{i,j}\ .
\label{varit}
\eea
Because each of $m$ iterations is statistically independent,
their separate results can
be combined into a single best answer and its estimated variance through
\bea
\bar{I} &=& {\sum_{i=1}^{m} I_i/\sigma_i^2 \over
\sum_{i=1}^{m}1/\sigma_i^2}\ ,\\
\bar{\sigma}^2 &=&  \left(\sum_{i=1}^{m}{1\over \sigma_i^2}\right)^{-1}\ ,
\eea
with the $\chi^2$ per degree of freedom given by
\beq
\chi^2/dof = \frac{1}{m-1}\sum_{i=1}^{m}
\frac{(I_i-\bar{I})^2}{\sigma_i^2}\ .
\eeq
When the algorithm is working properly, $\chi^2/dof$ should not
be much greater than one, since $(I_i-\bar{I})^2 \sim {\cal O}(\sigma_i^2)$.
Otherwise, different iterations are not consistent with each other.

\section{Parallelism considerations}
\label{mihne}

As mentioned earlier, the basic idea of
performing multidimensional Monte Carlo integration using many
computers is to find a scheme of dividing one  large calculation into
many small pieces, which can be handled separately and
in parallel. Because of that, the most natural framework for the
problem at hand is the so called {\em master/slave} model. In this
model the {\em master} program spawns the {\em slave} tasks and
distributes the different parts of the calculation to the
different slave processes.
These processes do their share of work, and send the results back to the
master program which combines them together.

The most important problem which one has to solve here is how to
divide the calculation between the slave tasks, while making sure
that the final result returned by the parallel algorithm does not depend on
factors such as the speed of different computers in the network,
the number of slave processes used for the calculation, etc.

The essential ingredient of our approach is that all parallel
tasks generate the same list of random numbers.
That is not
difficult to accomplish because all tasks use the same random number
generator, whose initial state is furnished to them by the master
program.
There are several good reasons for using this method:
\begin{itemize}
\item[(i)]
Reproducibility of the parallel algorithm can be easily achieved,
regardless of the number of parallel processes participating in
the calculation.
\item[(ii)]
Possibility of reproducing any part of the calculation,
which is important in case of possible failures of one or more
computers in the network.
\item[(iii)]
Low master-slave communication cost.
\end{itemize}
All of the above points will be  discussed in more details below.

\subsection{Parallel implementation of the VEGAS algorithm}

Since the number of available computers varies from situation to situation,
for parallelizing VEGAS we find it convenient to choose one
of the integration dimensions.
At the beginning of each iteration, the master program has to divide
the integration region in that
dimension into $n$
parts,\footnote{In this section
$x$ and $y$ are always coordinates along the integration dimension
used for parallelizing VEGAS. Also note that all coordinates are scaled:
if we have $z_L$ and
$z_U$ as the actual lower and upper boundaries of integration, then
the actual integration coordinate $z$ corresponds to
$x= (z -z_L)/(z_U-z_L)$, which ranges from 0 to 1.}
\beq
0 = y_0 < y_1 < \ldots < y_n = 1\ .
\eeq
Each subregion $\Delta y_i = y_i - y_{i-1}$ in the {\em task grid}
belongs to one parallel process. Note that the task grid
is different from the VEGAS' {\em sampling grid}, which divides
the same region into $K$ subdivisions,
\beq
0 = x_0 < x_1 < \ldots < x_K = 1\ ,
\eeq
with $\Delta x_k = x_k - x_{k-1}$.

In cases where only importance sampling
is used, the task $i$ has to evaluate the integrand only
if the random point
happens to fall within its one-dimensional subregion $\Delta y_i$.
In this way, all tasks accumulate
results for the entire integration volume. For stratified sampling
technique, which involves dividing
integration region into $M$ disjoint subvolumes, this strategy
would not be the most efficient one, since it would require keeping
track of results in all subvolumes. For large $M$ this would imply lots of
additional storage space in both master and slave programs, and also
a large overhead in the master-slave communication.\footnote{To
illustrate that, consider an example of $5$-dimensional integration
with requested $10^6$ function evaluations per iteration.
If stratified sampling were used,
the VEGAS algorithm would divide integration volume into approximately
$3.7\times 10^5$ subvolumes, and in each of them it would perform
a two-point Monte Carlo integration. Assuming the double precision arithmetic,
storing $S^{(1)}$ and $S^{(2)}$ for each subvolume would require
about $6$ megabytes of data, which would have to be passed by the
slave tasks to the
master program in each iteration.
}
Therefore, for stratified sampling it is more efficient to let one
task accumulate
all results within a given subvolume. This can be accomplished
because all parallel tasks generate the same list of random numbers.
Given its one-dimensional subregion boundaries,
once the task samples the first point for
integration in one particular
subvolume, it decides whether to accumulate results
in that subvolume, or to simply generate $N$ random points
without doing anything.
In other words, if the first point sampled in one particular subvolume
happened  to be in the subregion $\Delta y_i$ of the task grid,
then that subvolume belongs to the task $i$.
When this strategy is used, the work among
the parallel tasks is actually divided by subvolumes.

In either case, after it samples all of  $N\times M$
random points in one iteration,
the slave task sends accumulated results to the master program. Once
all results arrive, the master program
combines them to obtain the final iteration results for the integral and its
variance,
calculates the cumulative results for $\bar{I}$ and $\bar{\sigma}^2$,
and refines the sampling grid. Note that one of the advantages of this
approach is its minimal communication cost: the slave tasks receive all
necessary data (e.g., the sampling grid)
at the beginning of each iteration, and send the results back after
completing their share of work.

\subsection{Flexibility  issues}

Ideally, all tasks running in parallel would complete one particular iteration
at the same time, which would minimize their idle time, as well as
the total execution time of the parallel algorithm.
However, in a typical
network environment there are many factors which affect the performance
of the program running in parallel. For example, the calculation
may be affected by the different
computational speed of computers in the network, by the different
machine loads, etc. Furthermore, when the function being integrated
is concentrated in one particular region of space,
VEGAS quickly adjusts the sampling grid so that most of integrand
evaluations fall into that region. If that region happened to be
entirely within one subdivision $\Delta y_i$ of the task grid,
then the method we described above would give hardly any advantage over
the standard VEGAS algorithm,
since most of the work would have to be done by one task.
Because of these reasons,
it is essential that the  parallel algorithm has the ability
to adapt to a specific situation, i.e., to the given
network environment and to the function being integrated.

One possible solution to the above problems would be quite
simple:\footnote{We thank W.B. Kilgore for pointing out this possibility
to us.} assuming that we have $n$ parallel tasks participating
in the calculation, instead of dividing the task grid into
exactly $n$ subregions, we could divide it into $m$ parts, where
$m\gg n$.
After completing calculations in one particular subdivision
of the task grid, the slave task would
continue working on the next available one. In this way, the faster
processes would contribute
more to the calculation than the slower ones, and the optimal
work load could be achieved automatically. However, the problem
with this strategy is that it is associated with the
large cost of the communication  between the master
program and the slave tasks.

Because of that, in order to keep the communication cost low and still
achieve an optimal work distribution, we
propose to measure the time required
by the different tasks to complete their part of the calculation
in a given iteration, and to use that information
to distribute the work load for the next one. In other words, if,
for example,
the task $i$ takes longer than  others to complete its share of work,
its number of integrand
evaluations has to be decreased. This can be done simply
by adjusting the width of the subregion $\Delta y_i$
belonging to that task.

Denoting $a_i$ as
the time needed by the task $i$ for one integrand evaluation, and also
$b_i$ as the overhead time related to other necessary operations
(e.g., the random number generation),
the time  required by that  task  to complete one iteration  is given by
\beq
t_i = a_i N_i + b_i\ ,
\label{ti0}
\eeq
where $N_i$ is the number of integrand evaluations.
Constants $a_i$ and $b_i$ in Eq. (\ref{ti0}) are highly dependent on
the characteristics of the computer on which the task $i$ is running.
In order to determine them both, we would have to use the information
from the two successive iterations.
However, since our goal is to develop a scheme useful
for high  statistics integrations of computationally demanding functions,
we can safely assume that the parallel tasks spend most of their time
evaluating the integrand. This means that $a_i N_i \gg b_i$, and hence
\beq
t_i\simeq a_i N_i\ .
\label{t1}
\eeq
Let us also denote $t'_i$ as what would be the
optimal iteration completion time for the task $i$,
\beq
t'_i\simeq a_i N'_i\ ,
\label{t2}
\eeq
which is given in terms of the optimal
number of function evaluations $N'_i$.
As mentioned earlier, the perfect work load distribution among the parallel
tasks would be achieved if all of them finished their calculations
at the same time. Therefore, in the ideal case we would have
\beq
t'_i \simeq  \bar{t}' \ ,
\label{t3}
\eeq
where $\bar{t}'=\frac{1}{n}\sum_{j=1}^{n}t'_j$.
{}From Eqs. (\ref{t2}) and (\ref{t3}) we see that
$N'_i\propto 1/a_i$.
Using Eq. (\ref{t1}), and keeping in mind that
the total number  of function evaluations  in one iteration has
to be kept
constant ($N_T$), we put
\beq
N'_i =
N_T {N_i/t_i \over \sum_{j=1}^{n} N_j/t_j}\ ,
\label{ni}
\eeq
which satisfies the requirement
\beq
\sum_{i=1}^{n}N'_i = N_T\ .
\label{constraint}
\eeq
Even though the above derivation was rather heuristic, the final
expression for $N'_i$ is exact. It can be obtained in a more
rigorous way
using the method of Lagrange multipliers, by minimizing
the function
\beq
f(N'_1,N'_2, \ldots, N'_n) = \sum_{i=1}^{n}(\bar{t}'-t'_i)^2\ ,
\eeq
subject to the constraint given in Eq. (\ref{constraint}).

Eq. (\ref{ni})  allows the master program to use
the information from the previous iteration to
determine the optimal number of integrand evaluations
for the task $i$ in the next one.
Since the task work load can be adjusted by changing the width
of subdivisions
in the task grid, we still have to relate $N'_i$ to
$\Delta y_i$.
In the VEGAS algorithm, the probability of a random
point being generated within the $k$-th subdivision
of the sampling grid is given by \cite{lepage}
\beq
p(x) = \frac{1}{K\Delta x_k}\ ,\ \ \ \ \
{\rm for\ \ } x_{k-1} \leq x < x_k\ .
\label{px}
\eeq
Using this formula
it is not difficult to show that the expected number
of integrand evaluations for an arbitrary region between $x$ and $x'$
is given by
\bea
\tilde{N}(x,x') &=& N_T \int_{x}^{x'} dx\; p(x) \nonumber \\
&=& \frac{N_T}{K}
\left(k'-k + \frac{ x'-x_{k'-1}}{x_{k'}-x_{k'-1}}
- \frac{ x-x_{k-1}}{x_{k}-x_{k-1}}\right)\ ,
\label{nxx}
\eea
where we have assumed
$x_{k-1} \leq x < x_k $ and $x_{k'-1} \leq x' < x_{k'} $.
The above expression can be used to
find the optimal subdivisions $y_i$ of the task grid,
which are determined by
the relation
\beq
N'_i = \tilde{N}(y_{i-1}, y_i)\ .
\label{ni2}
\eeq
Given that $y_{i-1}$ is known ($y_0=0$), we
can solve  this equation for $y_i$,
\beq
y_i = x_{k'-1} + (x_{k'}-x_{k'-1})\left(\frac{K}{N_T}N'_i+k-k'+
\frac{ y_{i-1}-x_{k-1}}{x_{k}-x_{k-1}}\right)\ ,
\label{yi}
\eeq
and hence obtain the optimal task grid.
In Eq. (\ref{yi})
$k$ and $k'$ are again defined so that $x_{k-1}\leq y_{i-1} < x_k$
and $x_{k'-1}\leq y_{i} < x_{k'}$.

Since the distribution of the work load among the parallel tasks is
one of the most important ingredient of our approach, we summarize
it below:
\begin{enumerate}
\item
In each iteration the
slave tasks keep track of their actual number of integrand
evaluations, as well as of the time they require to complete the calculation.
\item
Using that information, the master program determines from Eq. (\ref{ni})
what would be the ideal number of function evaluations for each slave task.
\item
After computing the new sampling grid for the VEGAS algorithm,
the master program calculates the new task grid iteratively
using Eq. (\ref{yi}). Note that the boundary
conditions $y_0=0$ and $y_n=1$ have to be satisfied.
\end{enumerate}
In this way, after each iteration our algorithm adjusts
the task work load to achieve the best possible
performance in a given situation, while keeping low
cost of the communication between the master program and the slave tasks.

We should also mention that the above approach
can be easily generalized to allow the possibility
of dividing the calculation into $m\times n$ parts
($m\geq 2$), so that each of $n$ parallel tasks would work on
$m$ subregions  in one iteration. Although this would increase
the master-slave communication cost, as well as
the overhead time the slave tasks require in each iteration, it
would also shorten the time between the two successive task grid adjustments,
which  may be useful in an environment where the different machine
loads change rapidly.

\subsection{Robustness of the algorithm}

Another problem which has to be considered here is a possibility
that occasionally one or more parallel tasks may fail during the
calculation.\footnote{This can happen due to the lost network connection,
failure of a particular computer in the network, etc.}
Unless the algorithm has the ability to detect such an event,
and also to recalculate the lost results, the task failure
would require repeating the entire calculation.

In the scheme we are proposing in this paper, after distributing
various parts of the calculation to the parallel processes,
the master program
waits for all of them to complete their share of work and send
the results back. However, if none of the
results arrive after a certain amount of time, the master program
has to verify the current state of all parallel tasks.
In case that one or more tasks had failed, it
has to divide the lost parts of the calculation among the remaining
processes.

Even though the above strategy looks simple,
there are many details which have to be taken care of
in case of the task failure during the calculation. Nevertheless, since
the algorithm described in this section is ideally suited
for a recursive implementation of the work distribution,
it also allows for an efficient way of dealing with the task loss.

\section{The AMCI package}
\label{amci}

We have incorporated the general scheme described in the previous section
into a new Advanced Monte Carlo Integration
package.\footnote{The AMCI package can be obtained by sending an e-mail
to the author at {\bf veseli@fnal.gov}.} Besides
relying on the VEGAS algorithm \cite{lepage}, as well as on the long period
$(> 2\times 10^{18})$ random number
generator developed by P. L'Ecuyer \cite{lecuyer},
the package  also uses the standard set of communication
routines in the PVM software system \cite{pvm}.\footnote{The latest
version of the PVM software can be obtained by anonymous ftp
to {\bf netlib2.cs.utk.edu}, or from WWW by using the address
{\bf http://www.netlib.org/pvm3/index.html}.}

Since the scheme presented in Section \ref{mihne} is based
on the master/slave model, the AMCI package has two main parts:
the master and the slave subroutines,
each accompanied with functions taking care of the master-slave
communication via message-passing.
All of the AMCI functions, except the user-related ones, are placed into
several libraries which have to be linked with the driver
program. The package is written in the ANSI C programming language
(with the Fortran interface provided),
so that it should compile easily on all platforms which are also supported
by the PVM software system \cite{pvm}.\footnote{In
the near future, we hope to develop the
{\em Message Passing Interface} (MPI) \cite{mpi} version of the package.}

The most important
features of the package are as follows:
\begin{enumerate}
\item
Given the same seed for the random number generator,
the AMCI master routine  {\em always} returns the same answer as the ordinary
VEGAS algorithm (with or without stratified sampling),
regardless of the number of parallel tasks used for the
calculation.
\item
All of the useful features of the original VEGAS program are also
built into the AMCI package. For example,  the master
routine can be called again after initial preconditioning of the
sampling grid. There is also a possibility of computing
any number of arbitrary distributions of the sort
\beq
\frac{dI}{dy} = \int_{V} d\bfx\, f(\bfx)\delta(y-g(\bfx))\ ,
\eeq
with
\beq
I = \int dy \frac{dI}{dy}\ .
\eeq
\item
AMCI is flexible enough to adapt  to  specific conditions in the given
network environment, and also  to the particular function being integrated.
This property significantly increases the efficiency of the package.
For example, the master routine can be initially called  with
only a small number of integrand evaluations in a single iteration.
Even though all results obtained in that call would be discarded, this
procedure would allow AMCI to quickly optimize the task grid for the given
configuration of computers.
\item
AMCI has built in means of detecting  a possible task failure
and reproducing the lost parts of the calculation in an efficient way.
Because of that, the master program
is guaranteed to complete the calculation as long as at least one
slave task is running.
\item
The package is easy to use, and requires no knowledge of
parallel programming techniques.\footnote{After
the PVM software has been properly installed,
using the AMCI master subroutine is no more difficult than
using any of the standard subroutines from \cite{press}.}
\end{enumerate}
The last characteristic of the AMCI package is extremely important,
since it allows a typical user to benefit from distributed
computing, without becoming an expert in that area.

\section{Examples and performance analysis}
\label{perf}

\begin{figure}[p]
\vspace{12.0cm}
\includegraphics{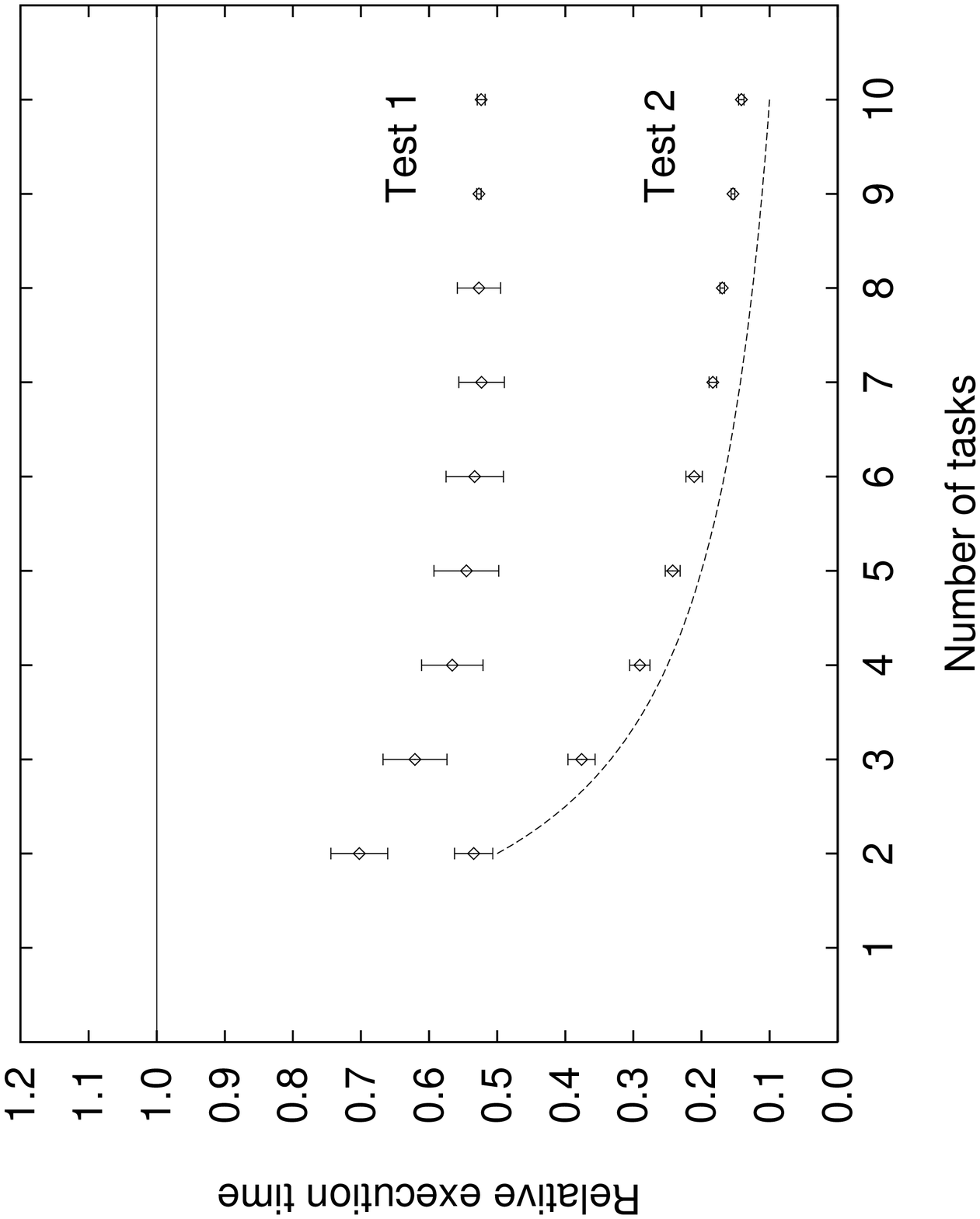}
\caption{AMCI execution time scaled with respect
to the execution time of the standard VEGAS program,
$\bT^{(n)}/T^{(1)}$, and
shown as the function of the number of parallel tasks  used for the
calculation. The actual
VEGAS execution times were 400  and 4366 seconds for tests 1 and 2,
respectively.
The dashed line denotes the ideal case, for which $\bT^{(n)}/T^{(1)}=1/n$.
}
\label{relative_time}
\end{figure}

\begin{figure}[p]
\vspace{12.0cm}
\includegraphics{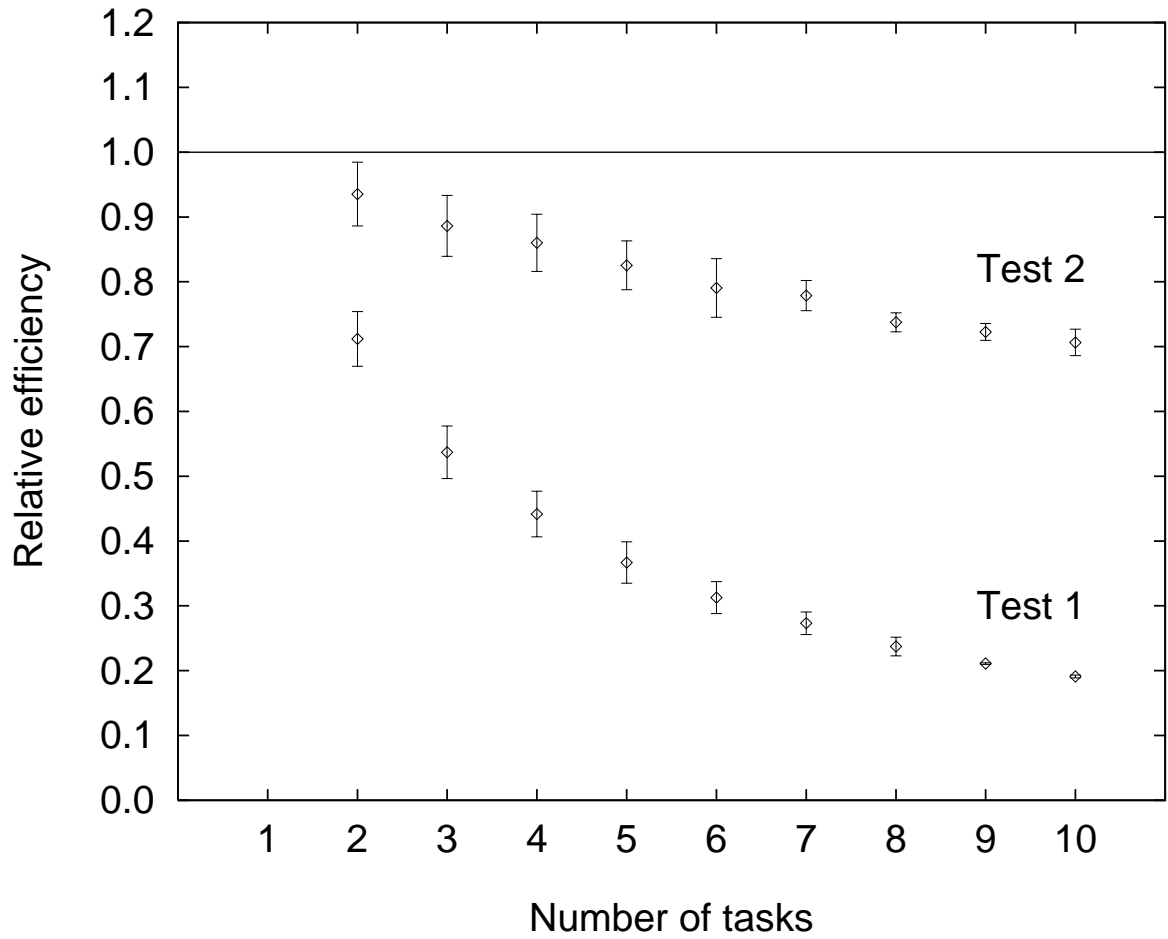}
\caption{AMCI efficiency for tests 1 and 2 scaled with respect
to the number of tasks participating in the calculation,
$T^{(1)}/n\bT^{(n)}$.
}
\label{efficiency}
\end{figure}

For comparison of the AMCI performance to that of
the standard VEGAS program,
we considered integration of a spherically symmetric
Gaussian placed in the center of the integration region,\footnote{Note
that the same test function was also used in \cite{lepage}.}
\beq
I_d = \left(\frac{1}{a \pi^{1/2}}\right)^d \int_0^1 d^d x\,
\exp\left(-\sum_{i=1}^{d}{(x_i-\frac{1}{2})^2 \over a^2} \right)\ ,
\label{test1}
\eeq
with $a = 0.1$.
As our PVM configuration we used $25$ $NeXT$ workstations
in the Fermilab Theory Group cluster. Most of those
machines  were equipped with 33 MHz processor, but some of them
had  25 MHz CPU's.
For each integration with $n$  requested parallel tasks ($2\leq n\leq 10$),
the PVM resource manager would decide which $n$
workstations would be used for the calculation.
In this way, we minimized  effects
of various factors, such as the speed of different computers,
different machine loads, etc. In order to further improve our
performance analysis, and to estimate the statistical errors, for
each $n$ we performed 10 independent integrations,
which means that $n$ parallel tasks were always running
on the different combination of $n$ computers from the PVM configuration.
We denote $\bT^{(n)}$ as
the average time required by the AMCI master routine to complete the
calculation using $n$ slave tasks.
On the other hand, the standard VEGAS program
was executed on all machines from the PVM configuration, and the
shortest execution time, denoted by $T^{(1)}$, was used for comparison
with $\bT^{(n)}$.
In Figures \ref{relative_time}  and \ref{efficiency} we show results for the
relative execution time  $\bT^{(n)}/T^{(1)}$, and for the relative efficiency
$T^{(1)}/n\bT^{(n)}$,  which were obtained in the two tests that were
performed.
In both figures we also show
the corresponding statistical errors.

The test   1  consisted of calculating the above integral in $d=5$ dimensions,
with about $10^5$ function evaluations in each of 10 iterations.
Even though the integrand
was relatively simple, with three tasks AMCI
has reduced the total execution  time to about 2/3 of the
time required by the standard VEGAS program (see Figure \ref{relative_time}).
However, addition of new  tasks
after $n = 4$ did not help significantly in terms of improving the
performance, which was still far from the ideal case
(shown with the dashed line). The reason for that is the simplicity
of the function being integrated:
in this particular case the execution time of the ordinary VEGAS
program was 400 seconds,
out of which about 40\% (160 seconds) was  used
for the random number generation.
Under the circumstances such as those, in which
the condition $a_i N_i\gg b_i$
is not satisfied and Eqs. (\ref{t1}) and (\ref{t2}) are
not valid,  the parallel algorithm
gets saturated with a small number of processes. Because of that
its efficiency as a function of the number of
processors participating in the calculation decreases rapidly,
which is illustrated in
Figure \ref{efficiency}.

In order to show how  the AMCI  performance with
respect to the standard VEGAS program improves as calculations
become more demanding,  for the test  2 we have artificially
slowed down the implementation of the
integrand function from Eq. (\ref{test1}). As a result, $T^{(1)}$
has been increased  by about a factor of 10,
from 400 to 4366 seconds, so that
the random number generation in this case
used less than 4\% of the total VEGAS execution time.
As shown in Figure \ref{relative_time},  the test 2 results for
$\bT^{(n)}/T^{(1)}$
follow the ideal $1/n$
curve much more closely than before, and statistical errors are
also reduced.
Consequently, the results for
the relative efficiency of the parallel algorithm are
significantly better than those obtained in the test 1
(see Figure \ref{efficiency}).\footnote{For the results
shown in Figures \ref{relative_time} and \ref{efficiency} one
has to  bear in mind that the {\em average} AMCI execution times
were compared to the {\em shortest} VEGAS execution time for
all of the machines from the PVM configuration,
and that not all of these computers were equally fast.
}

\begin{figure}[p]
\vspace{12.0cm}
\includegraphics{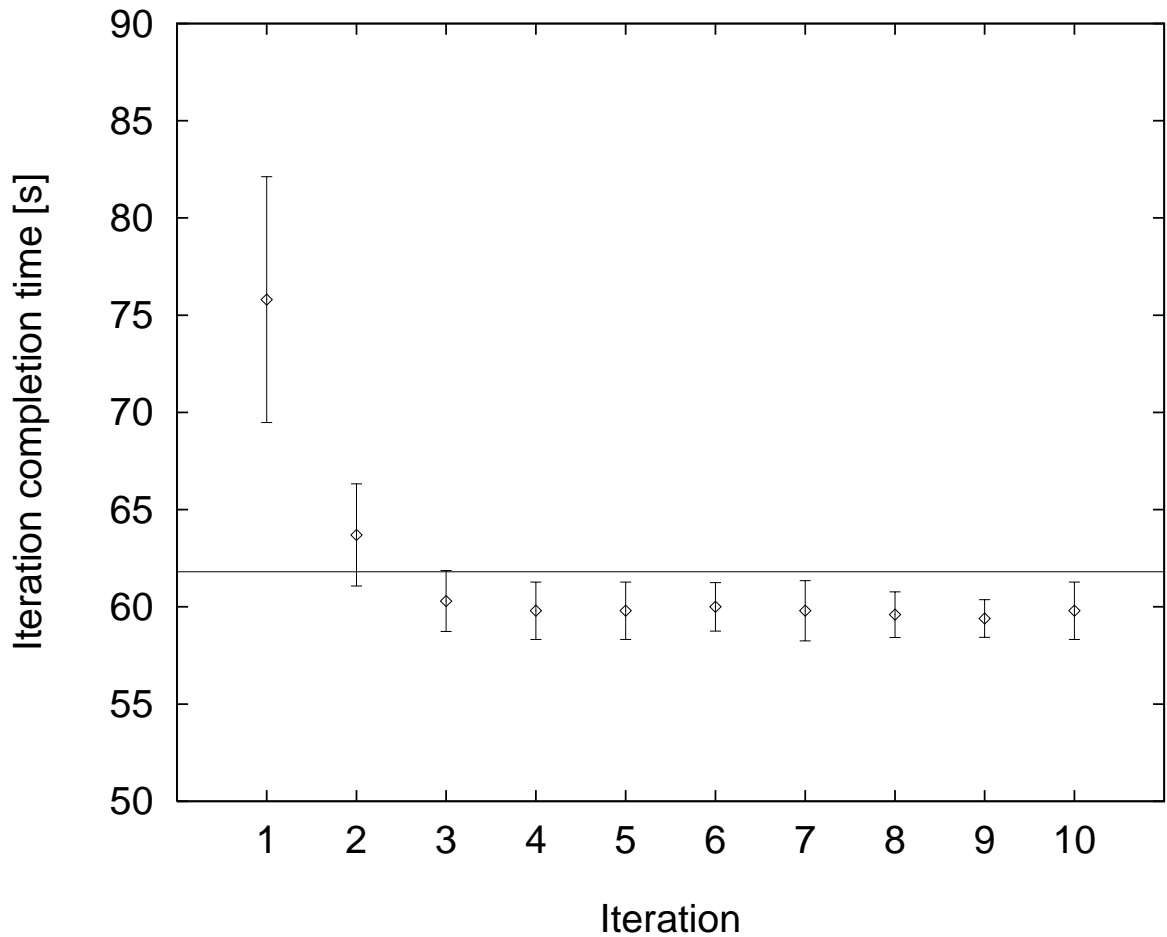}
\caption{The test 2 time required by AMCI running with $10$ parallel tasks
to complete the different iterations. The full line shows the average
iteration completion time. For comparison, the standard VEGAS
program required about 437 seconds for each iteration on the
fastest machine in the PVM configuration.
}
\label{test2_run10}
\end{figure}

\begin{figure}[p]
\vspace{12.0cm}
\includegraphics{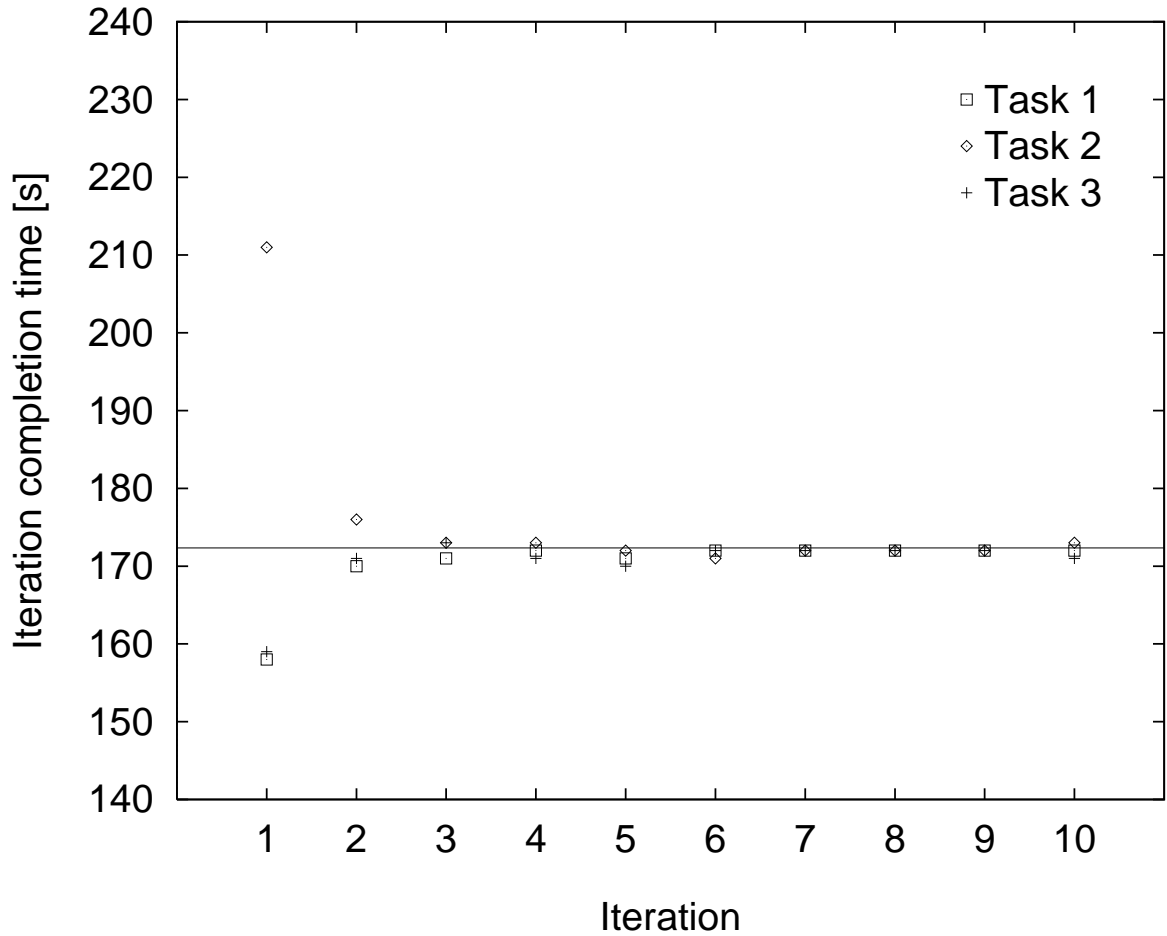}
\caption{The time required for the different tasks
to complete their  parts  of the calculation in the different iterations.
These results correspond to one of the test 2 runs with three
parallel tasks.
The full line shows the average iteration completion time.
For comparison, the standard VEGAS
program required about 437 seconds for each iteration on the
fastest machine in the PVM configuration.
}
\label{test2_run3}
\end{figure}

\begin{figure}[p]
\vspace{12.0cm}
\includegraphics{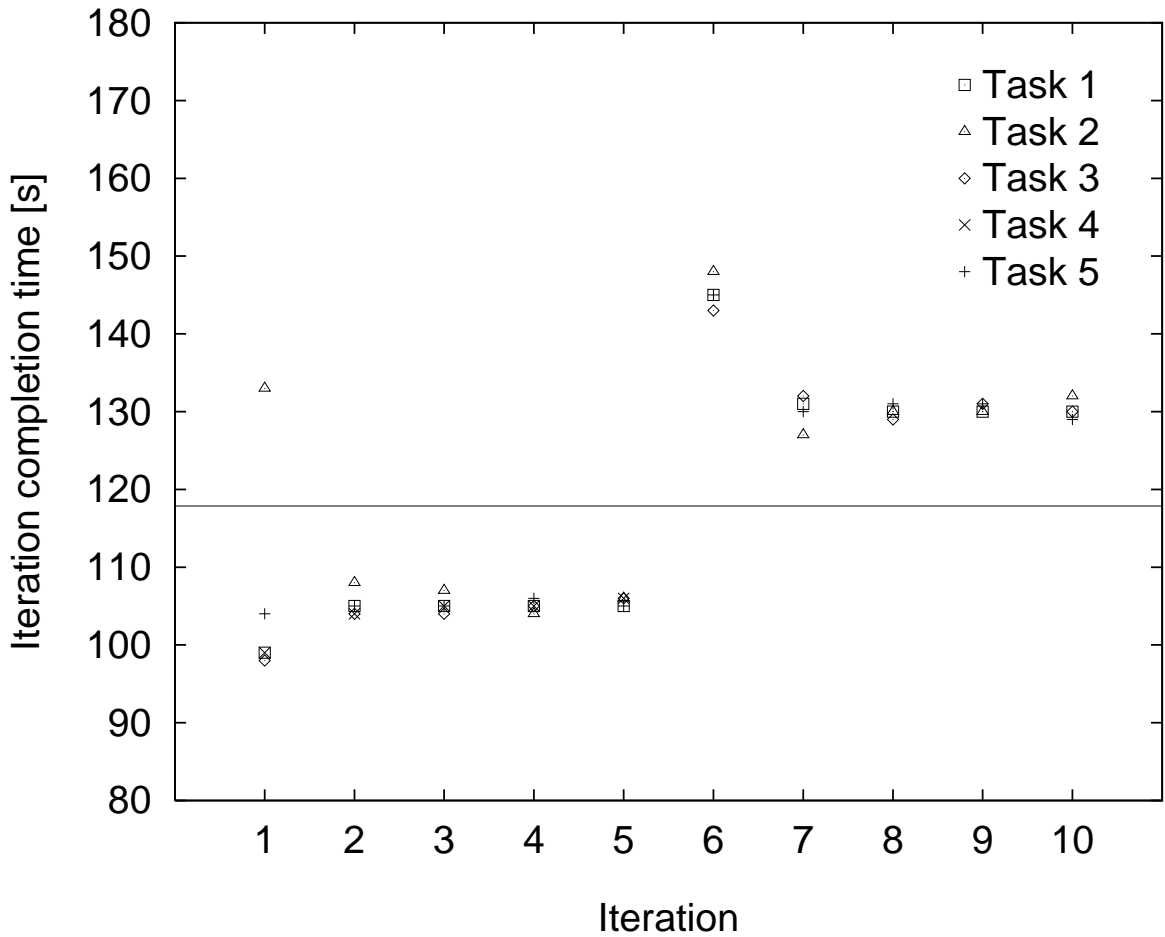}
\caption{The time required for the different tasks
to complete the different iterations
in the situation in which one of the tasks had failed.
To obtain these results, we repeated one of the test 2 runs by
starting with five parallel processes, and then removing
one of the computers from the PVM configuration. This caused failure
of the task 4 during the $6$-th iteration.
The average iteration completion time is shown with the full line.
For comparison, the standard VEGAS
program required about 437 seconds for each iteration on the
fastest machine in the PVM configuration.
}
\label{test3_run5}
\end{figure}

Figures \ref{test2_run10}, \ref{test2_run3} and \ref{test3_run5} are meant to
illustrate how the algorithm described in this paper
actually works, and how it behaves in various situations.
Figure \ref{test2_run10} shows the
average test 2 times, together with their respective statistical errors,
that were required by AMCI running with 10 parallel tasks to complete
the different iterations.
The longest time was needed for the first iteration, when
all tasks had to perform equal amounts of work. After the necessary
information about the different tasks was obtained in the first iteration,
the master subroutine quickly optimized the task grid
for the given configuration of computers.
We again point out that
for high statistics calculations better performance in the
first iteration can be achieved if the master subroutine is
initially called  with a small
number of integrand evaluations in a single iteration. This
would allow for the fast optimization of the task grid, and for the much
more efficient subsequent  calls with higher statistics.

Figure \ref{test2_run3} shows iteration completion times
for one of the test 2 runs with three parallel tasks. As one of the
computers used for that particular calculation was slower than
the other two, the task 2 took considerably longer time than tasks 1 and 3
to complete the first iteration. Again, this was accounted
for in subsequent iterations by optimizing the  work load for the
different tasks.

Figure \ref{test3_run5} illustrates the
behavior of the algorithm  in cases in which one of the
tasks fails during the calculation. In order to simulate that,
we repeated one of the test 2 calculations by starting with five
parallel tasks, and then removing  one of the
computers  from the
PVM configuration. This
caused  failure of the task 4  during the $6$-th iteration. Because
of that,
part  of the  calculation belonging to the task 4  had to be
divided among the remaining tasks.
After iteration 6 was
completed,  AMCI adjusted  to the new
situation, and the remaining tasks were again given the optimal work load.
Note that the time required for completing iteration 6 was only
about 15\% longer than the time needed for later iterations, which
shows that AMCI deals with task failures in an efficient way.

Finally, we briefly describe one real example from high energy
physics where AMCI  would be very useful:
theoretical description of the vector boson production at hadronic colliders.
This topic
is extremely important in view of
the precision measurements of the $W$ mass,
which may constrain parameters of the standard model (e.g., the Higgs mass).
At present, the state of the art
of the theory in the description of the vector boson production is based
on the resummation formalism of Collins, Soper, and Sterman \cite{CSS},
which involves
an inverse Fourier transform
of the cross section from the impact parameter  space to the
transverse momentum space.
Because of the oscillatory nature of the integrand in that Fourier transform,
the resummation calculations in the
impact parameter space are enormously difficult and lengthy.
For example, the program developed for the description
of the $W$ and $Z$ production \cite{ERV}, which is based on the
standard VEGAS algorithm, requires
more than 20 hours  on
an IBM $RS6000$ workstation to complete one calculation
with a very modest statistics of about $10^5$ (total) integrand
evaluations in the transverse momentum range from 0 to 50 GeV.
One should note here that the experimental analyses usually require
order(s) of magnitude higher statistics.

To make things even worse, the resummation formalism
also involves  several unknown
parameters, which have to be extracted from the experimental
vector boson transverse momentum distributions. In order to find the
best fit to the data,
calculations such as the one mentioned above
have to be repeated many times, once
for each different set of the non-perturbative parameters.

Even though the above numbers are just rough estimates,
they illustrate the fact that the theoretical description
of the vector boson production involves computationally extremely demanding
calculations, which  take a very long time with the standard
VEGAS program. On the other hand,  given $n$ equally fast computers,
the AMCI package would reduce
the VEGAS execution time by almost a factor of $1/n$, thus making
these calculations much more accessible.

\section{Conclusions}
\label{conc}

In this paper we have developed
a new parallel multidimensional
integration scheme,  suitable for use in a heterogeneous
network environment. This scheme, based on the well-known
adaptive Monte Carlo method (the VEGAS algorithm), is incorporated
into a new integration package (AMCI), which employs the standard
set of the message-passing routines in the PVM software system.
We have compared the AMCI performance with that of the
ordinary VEGAS program, and found that the new package is
significantly faster in cases involving high statistics
integrations of
computationally demanding functions.

\begin{center}
ACKNOWLEDGMENTS
\end{center}
The author owes a great deal of thanks to D. Donjerkovi\'{c} for
his comments and suggestions. He
would also like to thank W.B. Kilgore, W.F. Long and T. Stelzer
for useful discussions, and to express gratitude for hospitality
extended to him during the visit to UW-Madison Phenomenology Institute,
where part of this work was completed.
This work was supported in part by the U.S. Department of Energy
under Contract No. DE-AC02-76CH03000.

\end{document}